
\magnification \magstep1
\raggedbottom
\openup 2\jot
\voffset6truemm
\headline={\ifnum\pageno=1\hfill\else
{\it SPACE-TIME COVARIANT FORM OF ASHTEKAR'S CONSTRAINTS}
\hfill \fi}
\rightline {March 1995, DSF preprint 95/7}
\centerline {\bf SPACE-TIME COVARIANT FORM OF}
\centerline {\bf ASHTEKAR'S CONSTRAINTS}
\vskip 0.3cm
\centerline {\bf Giampiero Esposito$^{1,2}$, Gabriele Gionti$^{3}$,
Cosimo Stornaiolo$^{1,2}$}
\vskip 0.3cm
\noindent
{\it ${ }^{1}$Istituto Nazionale di Fisica Nucleare,
Sezione di Napoli,
Mostra d'Oltremare Padiglione 20, 80125 Napoli, Italy;}
\vskip 0.3cm
\noindent
{\it ${ }^{2}$Dipartimento di Scienze Fisiche,
Mostra d'Oltremare Padiglione 19, 80125 Napoli, Italy;}
\vskip 0.3cm
\noindent
{\it ${ }^{3}$Scuola Internazionale Superiore di Studi Avanzati,
Via Beirut 2-4, 34013 Trieste, Italy.}
\vskip 0.3cm
\noindent
{\bf Summary.} - The Lagrangian formulation of classical field
theories and in particular general relativity leads to a
coordinate-free, fully covariant analysis of these constrained
systems. This paper applies multisymplectic techniques to
obtain the analysis of Palatini and self-dual
gravity theories as constrained systems,
which have been studied so far in the Hamiltonian formalism.
The constraint equations are derived while paying attention
to boundary terms, and the Hamiltonian constraint turns out to
be linear in the multimomenta. The equivalence with
Ashtekar's formalism is also established.
The whole constraint analysis, however,
remains covariant in that the multimomentum
map is evaluated on {\it any} spacelike hypersurface.
This study is motivated by the non-perturbative
quantization program of general relativity.
\vskip 5cm
\leftline {PACS 04.20.Cv, 04.60.Ds}
\vskip 100cm
\leftline {\bf 1. - Introduction.}
\vskip 1cm
When Dirac developed his approach to constrained Hamiltonian
systems with the corresponding quantization program, he
emphasized that the Hamiltonian formalism is always necessary to
quantize a field theory with constraints. Remarkably, he provided
a well-defined (though not unique) prescription to define a
Hamiltonian function on the whole phase space, a classification
of constraints in terms of their Poisson brackets which is
immediately relevant for quantization, and an approach to
quantum electrodynamics which does not rely on Feynman diagrams
and renormalization theory [1-5]. However, any attempt to combine
Dirac's quantization of first-class constrained Hamiltonian
systems with the Arnowitt-Deser-Misner geometrodynamical
framework for canonical gravity faces very severe technical problems.
In other words, the occurrence of the scalar curvature of the
spacelike three-surfaces and the product of the three-momenta in the
Hamiltonian constraint make it impossible to find exact solutions
of the corresponding Wheeler-De Witt equation, as well as
interpret the (as yet unknown) physical states of the
quantum theory within the geometrodynamical framework [5,6].

More recently, the work by Ashtekar, Rovelli, Smolin and
their collaborators on connection dynamics and loop variables
has made it possible to cast the constraint equations of
general relativity in polynomial form, and then find a
large class of solutions to the
quantum version of constraints [7-14].
However, the quantum theory via the Rovelli-Smolin transform still
suffers from severe mathematical problems in 3+1 space-time
dimensions [15], and there appear to be reasons for studying
non-perturbative quantum gravity also from a Lagrangian, rather
than Hamiltonian, point of view (see below).
The aim of this paper is therefore
to provide a multisymplectic, Lagrangian framework for general
relativity [16-18],
to complement the present attempts to quantize
general relativity in a non-perturbative way. The motivations
of our analysis are as follows.

(i) In the case of field theories, there is not a unique
prescription for taking duals, on passing to the Hamiltonian
formalism. For example, algebraic and topological duals are
different. In turn, this may lead to inequivalent quantum
theories.

(ii) The 3+1 split of the Lorentzian space-time geometry, with the
corresponding $\Sigma \times R$ topology, appears to violate
the manifestly covariant nature of general relativity, as well
as rely on a very restrictive assumption on the
topology [19].

(iii) In the Lagrangian formalism, explicit covariance is
instead recovered. The first constraints one actually evaluates
correspond to the secondary first-class constraints of the
Hamiltonian formalism. At least at a classical level, the
Lagrangian theory of constrained systems is by now a
rich branch of modern mathematical physics [4,20,21],
although the majority of general relativists are more
familiar with the Hamiltonian framework.

(iv) In the ADM formalism [5,6], the invariance group is not
the whole diffeomorphism group, but a subgroup given by the
Cartesian product of diffeomorphisms on the real line with
the diffeomorphism group on spacelike three-surfaces. By contrast,
in the Lagrangian approach, the invariance group of the
theory is the full diffeomorphism group of four-dimensional
Lorentzian space-time. This is the unique group responsible for the
occurrence of constraints in Einstein's general relativity,
on looking at symmetry properties of the action functional
under such a group.

(v) Jet-bundles theory provides a rigorous geometric
framework for the hyperbolic problems of classical field
theory and, in particular, general relativity. The
corresponding evaluation of constraints, on using tetrad
formalism, is elegant, very powerful, and well-suited
for any attempt to study general relativity as a field
theory with constraints.

(vi) Although the elliptic boundary-value problems of
Riemannian geometry and quantum gravity via Wick-rotated
path integrals enable one to get a better
understanding of different approaches to the quantization
of gauge fields and gravitation [5], the corresponding
perturbative theory is non-renormalizable. Hence the
background-field method should be complemented or replaced
by a radically different view of space-time theories at the
Planck length [22].
The Hamiltonian approach is the first step
of the non-perturbative quantization program, but
unfortunately it breaks covariance.

We have thus tried to develop a classical multisymplectic analysis
of Palatini and self-dual gravity theories which might
be applied to a non-perturbative formalism for quantum
gravity that preserves covariance.
For this purpose, sect. {\bf 2} presents a
derivation of Einstein's equations in tetrad form which
relies on multisymplectic-geometry techniques (cf. [23]).
Section {\bf 3} studies multimomenta, sect. {\bf 4} derives constraint
equations in multisymplectic formalism, and their reduction to
Ashtekar's form is obtained in sect. {\bf 5}. The preservation of
constraints is analyzed in sect. {\bf 6}. Self-dual gravity
is studied in sect. {\bf 7}.
Concluding remarks and open problems are presented in sect. {\bf 8}.
Relevant mathematical background is described in the appendix.
\vskip 1cm
\leftline {\bf 2. - Multisymplectic form of Einstein's equations.}
\vskip 1cm
Our analysis begins by studying the Palatini action $S_{P}$
of (Lorentzian) general relativity. This is a real-valued
functional of the tetrad $e_{I}^{a}$ and the connection
one-form $\omega_{a}^{\; \; IJ}$, taking values in the Lorentz
Lie-algebra, and given by
$$
S_{P} \equiv
{1\over 2}\int_{M}d^{4}x \; e \; e_{\; I}^{a} \; e_{\; J}^{b}
\; \Omega_{ab}^{\; \; \; IJ} \; .
\eqno (2.1)
$$
With our notation, $e \equiv \sqrt{-g}$ is the square root of
the determinant of the space-time four-metric, and
$$
\Omega_{ab}^{\; \; \; IJ} \equiv
\partial_{a}\omega_{b}^{\; \; IJ}
-\partial_{b}\omega_{a}^{\; \; IJ}
+\omega_{a \; \; \; K}^{\; \; I}
\; \omega_{b}^{\; \; KJ}
-\omega_{b \; \; \; K}^{\; \; I}
\; \omega_{a}^{\; \; KJ}
\eqno (2.2)
$$
is the curvature of the four-dimensional connection one-form
$\omega_{a}^{\; \; IJ}$. Moreover, $a,b$ are tangent-space
indices, whereas $I,J$ are the usual internal
indices [13].

In first-order theory, tetrad and connection are regarded as
independent variables. Since we are aiming to use
a Lagrangian version of first-order theory in terms of one-jet
bundles (see appendix),
it is useful to bear in mind that
in general relativity one takes a fibre bundle
whose base is space-time, and
whose fibres are isomorphic to the Cartesian product of
the space of Lorentzian four-metrics with the space of linear
connections. In the language of tetrads and connection one-forms
used in (2.1)-(2.2), one takes a fibre bundle $Y$ which,
in local coordinates, reads
$$
\Bigr(x^{a},e_{\; \; I}^{a},\omega_{a}^{\; \; IJ}\Bigr) \; .
$$
To obtain the corresponding one-jet bundle $J^{1}(Y)$ (see
appendix), one is thus led to consider the multivelocities
$V_{\; \; b I}^{a}$ corresponding to the tetrad, and
the multivelocities $W_{ab}^{\; \; \; IJ}$ corresponding to the
connection one-form. In local coordinates, our one-jet bundle
$J^{1}(Y)$ is therefore represented by
$$
\biggr(x^{a},e_{\; \; I}^{a},\omega_{a}^{\; \; IJ},
V_{\; \; b I}^{a},W_{ab}^{\; \; \; IJ}\biggr) \; .
$$
This leads to the Lagrangian
$$
L \equiv {e\over 2} \; e_{\; I}^{a} \; e_{\; J}^{b}
\Bigr(W_{ab}^{\; \; \; IJ}-W_{ba}^{\; \; \; IJ}
+[\omega_{a},\omega_{b}]^{IJ}\Bigr) \; ,
\eqno (2.3)
$$
and hence to the Cartan four-form on $J^{1}(Y)$ as
$$ \eqalignno{
\Theta_{L}&=\left(L-{\partial L \over \partial W_{ab}^{\; \; \; IJ}}
\; W_{ab}^{\; \; \; IJ}\right)d^{4}x
+{\partial L \over \partial W_{ab}^{\; \; \; IJ}}
\; d\omega_{a}^{\; \; IJ}\wedge d^{3}x_{b} \cr
&={e\over 2} \Bigr(e_{\; I}^{a} \; e_{\; J}^{b}
-e_{\; I}^{b} \; e_{\; J}^{a}\Bigr)
\Bigr[d\omega_{a}^{\; \; IJ} \wedge d^{3}x_{b}
+\omega_{a \; \; \; K}^{\; \; I}
\; \omega_{b}^{\; \; KJ} \; d^{4}x \Bigr] \; .
&(2.4)\cr}
$$
Note that $V_{\; \; b I}^{a}$ does not contribute,
since derivatives of the tetrad do not occur in this first-order
theory for vanishing torsion.
The corresponding multisymplectic five-form $\Omega_{L}$
is obtained by exterior differentiation of $\Theta_{L}$
as $\Omega_{L} \equiv d\Theta_{L}$.
To write down field equations, one now considers a vector field
$U$ tangent to $J^{1}(Y)$. It has the form
$$
U= U^{d}{\partial \over \partial x^{d}}
+U_{\; I}^{d} {\partial \over \partial e_{\; I}^{d}}
+U_{d}^{\; \; IJ}
{\partial \over \partial \omega_{d}^{\; \; IJ}}
+U_{\; \; bI}^{d}{\partial \over \partial V_{\; \; bI}^{d}}
+U_{dl}^{\; \; \; IJ}
{\partial \over \partial W_{dl}^{\; \; \; IJ}} \; .
\eqno (2.5)
$$
One then takes the contraction of $\Omega_{L}$ with $U$. The
pull-back of the resulting geometric object by means of the
tangent lift $j^{(1)}(\varphi)$
of sections of the fibre bundle $Y$, leads
to the Euler-Lagrange field equations.
In our case one obtains
$$ \eqalignno{
i_{U}\Omega_{L}&
=-{e\over 2}\Bigr(e_{\; I}^{a} \; e_{\; J}^{b}
-e_{\; I}^{b} \; e_{\; J}^{a}\Bigr)
e_{\; h}^{L} \;
de_{\; L}^{h} \wedge d\omega_{a}^{\; \; IJ} \wedge
d^{2}x_{bm} \; U^{m} \cr
&+{e\over 2}
\Bigr[(de_{\; I}^{a})e_{\; J}^{b}
+e_{\; I}^{a}(de_{\; J}^{b})-(de_{\; I}^{b})
e_{\; J}^{a}-e_{\; I}^{b}(de_{\; J}^{a})\Bigr]
\wedge d\omega_{a}^{\; \; IJ} \wedge d^{2}x_{bl} \; U^{l}\cr
&+{e\over 2}\Bigr(e_{\; I}^{a} \; e_{\; J}^{b}
-e_{\; I}^{b} \; e_{\; J}^{a}\Bigr)e_{\; h}^{L}
\Bigr(de_{\; L}^{h} \wedge d^{3}x_{m} \; U^{m}\Bigr)
\omega_{a \; \; \; K}^{\; \; I}
\; \omega_{b}^{\; \; KJ}\cr
&-{e\over 2} \Bigr[(de_{\; I}^{a})e_{\; J}^{b}
+e_{\; I}^{a}(de_{\; J}^{b})-(de_{\; I}^{b})
e_{\; J}^{a}-e_{\; I}^{b}(de_{\; J}^{a})\Bigr]
\omega_{a \; \; \; K}^{\; \; I}
\; \omega_{b}^{\; \; KJ} \wedge d^{3}x_{l} \; U^{l}\cr
&-{e\over 2}\Bigr(e_{\; I}^{a} \; e_{\; J}^{b}
-e_{\; I}^{b} \; e_{\; J}^{a}\Bigr)
\left[\Bigr(d\omega_{a \; \; \; K}^{\; \; I}\Bigr)
\omega_{b}^{\; \; KJ}+\omega_{a \; \; \; K}^{\; \; I}
\Bigr(d\omega_{b}^{\; \; KJ}\Bigr)\right]
\wedge d^{3}x_{l} \; U^{l} \cr
&+{e\over 2}\Bigr[U_{\; I}^{a} \; e_{\; J}^{b}
+U_{\; J}^{b} \; e_{\; I}^{a}
-U_{\; I}^{b} \; e_{\; J}^{a}
-U_{\; J}^{a} \; e_{\; I}^{b} \Bigr]
\biggr[d\omega_{a}^{\; \; IJ} \wedge d^{3}x_{b}
+\omega_{a \; \; \; K}^{\; \; I} \;
\omega_{b}^{\; \; KJ} \; d^{4}x \biggr] \cr
&-{e\over 2}\Bigr(e_{\; I}^{a} \; e_{\; J}^{b}
-e_{\; I}^{b} \; e_{\; J}^{a}\Bigr)
e_{\; h}^{L} \; U_{\; L}^{h}
\biggr[d\omega_{a}^{\; \; IJ} \wedge d^{3}x_{b}
+\omega_{a \; \; \; K}^{\; \; I} \;
\omega_{b}^{\; \; KJ} \; d^{4}x \biggr] \cr
&+{e\over 2}\Bigr(e_{\; I}^{a} \; e_{\; J}^{b}
-e_{\; I}^{b} \; e_{\; J}^{a}\Bigr)
U_{a}^{\; \; IJ} \; e_{\; h}^{L} \; de_{\; L}^{h}
\wedge d^{3}x_{b} \cr
&-{e\over 2} U_{a}^{\; \; IJ}
\Bigr[(de_{\; I}^{a})e_{\; J}^{b}
+e_{\; I}^{a}(de_{\; J}^{b})-(de_{\; I}^{b})
e_{\; J}^{a}-e_{\; I}^{b}(de_{\; J}^{a})\Bigr]
\wedge d^{3}x_{b} \cr
&+{e\over 2}\Bigr(e_{\; I}^{a} \; e_{\; J}^{b}
-e_{\; I}^{b} \; e_{\; J}^{a}\Bigr)
\Bigr(U_{a \; \; \; K}^{\; \; I} \; \omega_{b}^{\; \; KJ}
+\omega_{a \; \; \; K}^{\; \; I} \;
U_{b}^{\; \; KJ}\Bigr) d^{4}x \; .
&(2.6)\cr}
$$
Such a lengthy equation is indeed necessary, since it involves
many contributions which cannot be obvious to non-expert
readers, and their interpretation needs a careful thinking.

When one evaluates the pull-back of $i_{U}\Omega_{L}$ by
means of the tangent lift
$$
\Bigr(j^{(1)}(\varphi)\Bigr) \equiv
\biggr(x^{a}, e_{\; \; I}^{a}(x),\omega_{a}^{\; \; IJ}(x),
{\partial e_{\; \; I}^{a}\over \partial x^{b}}(x),
{\partial \omega_{a}^{\; \; IJ}\over \partial x^{b}}(x)
\biggr) \; ,
$$
the terms $de_{\; \; L}^{h}$ and $d\omega_{a}^{\; \; IJ}$
occurring in (2.6) take the forms
$$
de_{\; L}^{h}=e_{\; L,f}^{h} \; dx^{f} \; ,
\eqno (2.7)
$$
$$
d\omega_{a}^{\; \; IJ}=\omega_{a \; \; \; \; ,l}^{\; \; IJ}
\; dx^{l} \; .
\eqno (2.8)
$$
Thus, by using the identities
$$
dx^{f} \wedge d^{3}x_{l}=\delta_{l}^{f} \; d^{4}x \; ,
\eqno (2.9)
$$
$$
dx^{f}\wedge dx^{h} \wedge d^{2}x_{bc}
=\Bigr(\delta_{b}^{f} \; \delta_{c}^{h}
-\delta_{c}^{f} \; \delta_{b}^{h}\Bigr)
d^{4}x \; ,
\eqno (2.10)
$$
and defining
$$
\delta e_{\; L}^{h} \equiv U^{m} \; e_{\; L,m}^{h}
-U_{\; L}^{h} \; ,
\eqno (2.11)
$$
$$
\delta \omega_{a}^{\; \; IJ} \equiv
U^{m} \; \omega_{a \; \; \; \; ,m}^{\; \; IJ}
-U_{a}^{\; \; IJ} \; ,
\eqno (2.12)
$$
$$
D_{b}e_{\; I}^{a} \equiv e_{\; I,b}^{a}
+\omega_{b \; \; \; I}^{\; \; L} \; e_{\; L}^{a} \; ,
\eqno (2.13)
$$
the field equations
$$
\Bigr(j^{(1)}(\varphi)\Bigr)^{*}
\Bigr(i_{U}\Omega_{L}\Bigr)=0
\eqno (2.14)
$$
are found to take the form
$$
e \Bigr(\delta e_{\; L}^{h}\Bigr)G_{\; \; h}^{L}
+\Bigr(\delta \omega_{a}^{\; \; IJ}\Bigr)F_{\; \; IJ}^{a}=0 \; ,
\eqno (2.15)
$$
where
$$
G_{\; \; h}^{L} \equiv 2e_{\; I}^{b}
\left[\Omega_{bh}^{\; \; \; \; IL}
-{1\over 2}e_{\; J}^{d} \; e_{\; h}^{L} \;
\Omega_{bd}^{\; \; \; \; IJ}\right] \; ,
\eqno (2.16)
$$
$$
F_{\; IJ}^{a}
\equiv D_{b} \left[e\Bigr(
e_{\; I}^{b} \; e_{\; J}^{a}
-e_{\; I}^{a} \; e_{\; J}^{b}
\Bigr)\right] \; .
\eqno (2.17)
$$
Since $e_{\; \; L}^{h}$ and $\omega_{a}^{\; \; IJ}$ are
independent, Eq. (2.15) implies
$$
e_{\; \; I}^{b} \left[\Omega_{bh}^{\; \; \; \; IL}
-{1\over 2}e_{\; J}^{d} \; e_{\; h}^{L} \;
\Omega_{bd}^{\; \; \; \; IJ}\right]=0 \; ,
\eqno (2.18)
$$
$$
D_{b} \left[e\Bigr(e_{\; I}^{a} \;
e_{\; J}^{b}-e_{\; I}^{b} \; e_{\; J}^{a}\Bigr)\right]=0 \; .
\eqno (2.19)
$$
Eqs. (2.18) are the Einstein equations, while eqs. (2.19) express
a property of a connection which is completely determined
by the tetrad [13].

Note that in (2.11)-(2.12) the terms involving partial
derivatives of the field variables are the horizontal
part of the variation, and remaining
terms represent the vertical part of the variation.
Hence (2.11)-(2.12) have a very clear geometric meaning,
and they lead to a considerable simplification of a
lengthy calculation.
Moreover, (2.13) defines the covariant derivatives of
the tetrad. The connection $D$ is a Lorentz connection
which annihilates the Minkowskian metric $\eta_{IJ}$
on the internal space. The definition (2.13) is an
additional condition we are imposing, since the action
of $D$ on space-time indices is not defined
{\it a priori} [13].
\vskip 10cm
\leftline {\bf 3. - Multimomenta.}
\vskip 1cm
The analysis of constraint equations in sect. {\bf 4} makes it
necessary to describe some properties
of the multimomenta corresponding to the multivelocities
defined in sect. {\bf 2}. The
multimomenta of a field theory are defined as the derivatives
of the Lagrangian with respect to the multivelocities. In
general relativity, the multimomenta resulting from the
Lagrangian (2.3) are defined as the densities
$$
{\tilde p}_{\; \; \; IJ}^{ab}
\equiv 2 {\partial L \over \partial W_{ab}^{\; \; \; IJ}}=
e \; p_{\; \; \; IJ}^{ab} \; ,
\eqno (3.1)
$$
$$
{\widetilde \pi}_{a}^{\; \; b I} \equiv
2{\partial L \over \partial V_{\; \; bI}^{a}}=0 \; ,
\eqno (3.2)
$$
where the $p_{\; \; \; IJ}^{ab}$ are bivectors defined as
$$
p_{\; \; \; IJ}^{ab} \equiv e_{\; \; I}^{a} \;
e_{\; \; J}^{b}-e_{\; \; I}^{b} \; e_{\; \; J}^{a} \; .
\eqno (3.3)
$$
Of course, the multiplicative factor in (3.1)-(3.2) is
unessential, and is introduced for convenience. Note that
the multimomenta ${\widetilde \pi}_{a}^{\; \; bI}$ vanish, and
this reflects that torsion vanishes in general relativity.

{}From the definition (3.3), we note that
the bivectors $p_{\; \; \; IJ}^{ab}$
satisfy the following commutation relations:
$$
\Bigr[p^{ab}, p^{cd}\Bigr]_{IJ}=
p_{\; \; \; IJ}^{ac} \; g^{bd}
+ p_{\; \; \; IJ}^{bd} \; g^{ac}
- p_{\; \; \; IJ}^{ad}
\; g^{bc} - p_{\; \; \; IJ}^{bc} \; g^{ad} \; .
\eqno (3.4)
$$
In particular, on a spacelike hypersurface, Eq. (3.4) may
be used to derive the identity
$$
\Bigr[p^{i0}, p^{j0}\Bigr]_{IJ}
= p_{\; \; \; IJ}^{ij} \; g^{00}
- p_{\; \; \; IJ}^{i0}
\; g^{0j} - p_{\; \; \; IJ}^{0j} \; g^{i0} \; ,
\eqno (3.5)
$$
where the index $0$ refers to the time coordinate and
the indices $i,j$ refer to the spatial coordinates.
This property will be useful in sect. {\bf 5}.
The multimomenta correspond to
36 variables (6 for each bivector density),
therefore equation (3.5)
expresses the eighteen variables associated
to the ${\tilde p}_{\; \; \; IJ}^{ij}$ as
quadratic functions of the
${\tilde p}_{\; \; \; IJ}^{0i}$ bivector densities.
In the following sections the constraint equations are all
expressed in terms of the multimomenta.
\vskip 1cm
\leftline {\bf 4. - Constraints.}
\vskip 1cm
When one studies classical mechanics and classical field
theory one learns that, if the Lagrangian is invariant
under the action of a group, then by virtue of Noether's
theorem there exist functions which are constant along
solutions of the equations of motion. In the case of the
invariance under a gauge group or the diffeomorphism group
of general relativity, such first integrals always vanish
along solutions of the equations of motion. Hence the
first-class constraints of a field
theory result from Noether's theorem
through the action of the
gauge group or the group of space-time diffeomorphisms [23,24].
The multimomentum map
is the mathematical tool which enables one to describe these
properties of classical fields.
With the notation of Eqs. (A.1)-(A.2),
the multimomentum map on a section
of our jet-bundle $J^{1}(Y)$ is defined by the expression [23,24]
$$
\Bigr(j^{(1)}(\varphi)\Bigr)^*
\Bigr[{\cal FL}^*(J_\xi)\Bigr] \equiv
\left[{\partial L\over \partial {\varphi}_{\; \; \; \; ,a}^{(A)}}
\Bigr(\xi^{(A)}-
{\varphi}_{\; \; \; \; ,b}^{(A)} \;
\xi^{b}\Bigr)+L\xi^{a} \right] d^{3}x_{a} \; ,
\eqno (4.1)
$$
where $\xi^{(A)}$ is the variation along
the fibre and $\xi^{a}$ is the variation
along the base space. In terms of the variables
$e_{\; \; I}^{a}$ and $\omega_{a}^{\; \; IJ}$ the
multimomentum map $J$ becomes
$$
J(\xi)= \biggr[e{\partial L\over \partial e^a_{\;\;I,c}}
\Bigr(\xi_{I}^{a} -e^{a}_{\;\;I,b} \; \xi^{b}\Bigr)
+ e{\partial L \over \partial \omega_{a \; \; \; \; ,c}^{\; \; IJ}}
\Bigr(\xi^{IJ}_a - \omega_{a \; \; \; ,b}^{\; \; IJ}
\; \xi^{b}\Bigr)
+ {e \over 2} \; e_{\; \; I}^{a}
\; e_{\; \; J}^{b} \; \Omega_{ab}^{\; \; \; IJ}
\; \xi^{c}\biggr]  d^{3}x_{c}.
\eqno (4.2)
$$
In the Hamiltonian framework, setting to zero the integral of
the multimomentum map
$J_{\xi}$ on a spacelike hypersurface $\Sigma$
leads to the first-class constraints of the theory [23,24].
Moreover, in the case of null hypersurfaces,
second-class constraints also occur [25], and hence they cannot be
derived from the multimomentum map. On passing from
spacelike to null hypersurfaces, discontinuities occur
in the normal to the hypersurface and in the induced three-metric [25].
Hence there might be a covariant constraint analysis for any
spacelike hypersurface, and an independent, covariant constraint
analysis for any null hypersurface. [We are grateful to Luca
Lusanna for bringing this open problem to our attention]

To express our $J(\xi)$, from
$$
{\partial L\over \partial e^{a}_{\; \; I,c}}=0 \; ,
\eqno (4.3)
$$
and
$$
\xi^{\; \; IJ}_{a}
= -\xi^b_{\ ,a} \; \omega_{b}^{\ IJ}
+ (D_{a} \lambda)^{\ IJ} \; ,
\eqno (4.4)
$$
it follows (on using (3.3))
$$ \eqalignno{
I_{\Sigma}[\xi] &\equiv \int_{\Sigma}J(\xi)
= {1\over 2}\int_{\Sigma}
\biggr[e \; p_{\; \; \; IJ}^{ac} \Bigr(\xi^b_{,a}
\; \omega_{b}^{\; \; IJ} - (D_{a} \lambda)^{\ IJ}+
\omega_{a \; \; \; \; ,b}^{\; \; IJ} \; \xi^b\Bigr)\cr
&+ {1\over 2} e \; p_{\; \; \; IJ}^{ab} \; \Omega_{ab}^{\; \; \; IJ}
\; \xi^{c}\biggr] d^{3} x_{c} \; ,
&(4.5)\cr}
$$
where one has defined the covariant derivative with respect to the tetrad
indices (cf. (2.13))
$$
(D_{a} \lambda)^{\ IJ} \equiv
(\partial_{a}\lambda)^{\ IJ}+[\lambda,\omega_{a}]^{IJ} \; .
\eqno (4.6)
$$
Thus, taking the integral of the multimomentum map
on the spacelike hypersurface $\Sigma$,
integrating by parts for the term in $\lambda_{I}^{\ J}$,
and defining
$$
\sigma^{ac} \equiv {\tilde p}_{\; \; \; IJ}^{ac}
\; \lambda^{IJ} \; ,
\eqno (4.7)
$$
one obtains
$$ \eqalignno{
I_{\Sigma}[\xi]&=
{1\over 2}\int_{\Sigma} \lambda^{IJ}
(D_{a}{\tilde p}^{ac})_{IJ} \; d^{3}x_{c}
+ {1\over 2}\int_{\Sigma} \left[
({\tilde p}_{\; \; \; IJ}^{ac})
(L_{\xi}\omega)_{a}^{\ IJ}
+ {1\over 2} {\tilde p}_{\; \; \; IJ}^{ab}
\; \Omega_{ab}^{\; \; \; IJ} \; \xi^{c}\right]
d^{3}x_{c} \cr
&-{1\over 2} \int_{\Sigma}\partial_{a}\sigma^{ac}
\; d^{3}x_{c} \; .
&(4.8)\cr}
$$
Again from integration by parts in the second integral,
and defining
$$
\rho^{ac} \equiv {\tilde p}_{\; \; \; IJ}^{ac}
\; \omega_{b}^{\; \; IJ} \; \xi^{b} \; ,
\eqno (4.9)
$$
one finds
$$ \eqalignno{ I_{\Sigma}[\xi]&={1\over 2}
\int_{\Sigma} \left[
-\xi^{b}({\tilde p}_{\; \; \; IJ}^{ac})
\Bigr(\omega_{b \; \; \; \; ,a}^{\; \; IJ}
- \omega_{a \; \; \; \; ,b}^{\; \; IJ}\Bigr)
-\xi^{b} \; {\tilde p}_{\; \; \; IJ,a}^{ac}
\; \omega_{b}^{\; \; IJ}
+ {1\over 2} {\tilde p}_{\; \; \; IJ}^{ab}
\; \Omega_{ab}^{\; \; \; IJ} \; \xi^{c} \right]
d^{3}x_{c} \cr
&+{1\over 2}\int_{\Sigma}\partial_{a}\rho^{ac} d^{3}x_{c}
+{1\over 2} \int_{\Sigma} \lambda^{IJ}
(D_{a}{\tilde p}^{ac})_{IJ}\; d^{3}x_{c}
-{1\over 2}\int_{\Sigma}\partial_{a}\sigma^{ac} d^{3}x_{c} \; .
&(4.10)\cr}
$$
Using equation (2.19), and denoting by $I_{\Sigma}^{\rm ess}[\xi]$
the part of $I_{\Sigma}[\xi]$ not involving total divergences,
one can show that
$$ \eqalignno{
I_{\Sigma}^{\rm ess}[\xi]&={1\over 2}
\int_{\Sigma} \lambda^{IJ} (D_{a}{\tilde p}^{ac})_{IJ} \;
d^{3}x_{c} + {1\over 2} \int_{\Sigma}\left[-
({\tilde p}_{\; \; \; IJ}^{ac})\Omega_{ad}^{\; \; \; IJ}
+ {1\over 2} {\tilde p}_{\; \; \; IJ}^{ab}
\; \Omega_{ab}^{\; \; \; IJ} \; \delta_{d}^{c} \right]
\xi^{d} d^{3}x_{c} \cr
&={1\over 2}
\int_{\Sigma} \lambda^{IJ}
(D_{a}{\tilde p}^{ac})_{IJ} \; d^{3}x_{c}
- {1\over 2} \int_{\Sigma}
{\rm Tr} \left[{\tilde p}^{ac}\Omega_{ad}
- {1\over 2} {\tilde p}^{ab}\Omega_{ab}
\; \delta_{d}^{c} \right] \xi^{d}
d^{3}x_{c}  \; ,
&(4.11)\cr}
$$
where $\lambda^{IJ}$ and $\xi^{d}$ are independent
and arbitrary quantities. Then from
imposing $I_{\Sigma}[\xi]=0$, one finds two sets of constraints:
$$
\int_{\Sigma}\partial_{a}\sigma^{ac} \; d^{3}x_{c}
-\int_{\Sigma} \lambda^{IJ}
(D_{a}{\tilde p}^{ac})_{IJ}\;  d^{3}x_{c}=0 \; ,
\eqno (4.12)
$$
and
$$
\int_{\Sigma}\partial_{a}\rho^{ac} \; d^{3}x_{c}
-\int_{\Sigma} {\rm Tr} \left[{\tilde p}^{ac}\Omega_{ad}
- {1\over 2} {\tilde p}^{ab}\Omega_{ab}
\; \delta_{d}^{c}\right]
\xi^{d} \; d^{3}x_{c} =0 \; .
\eqno (4.13)
$$
Note that the total divergences appearing in (4.12)-(4.13)
lead to boundary terms evaluated on the
{\it two-surface} $\partial \Sigma$. Hence
sufficient conditions for the vanishing of such boundary
terms are as follows:
(i) $\Sigma$ has no boundary;
(ii) ${\tilde p}_{\; \; \; IJ}^{ab}$ vanishes at
$\partial \Sigma$;
(iii) $\lambda^{IJ}$ vanishes at $\partial \Sigma$
(in (4.12));
(iv) $\xi^{b}$ or $\omega_{b}^{\; \; IJ}$ vanishes at
$\partial \Sigma$ (in (4.13)).
Although these conditions are (rather) restrictive, from now on
we will always assume that (i) or (ii) or both (iii) and (iv)
hold. Hence (4.12)-(4.13) reduce to
$$
\int_{\Sigma} \lambda^{IJ}
(D_{a}{\tilde p}^{ac})_{IJ} \; d^{3}x_{c}=0 \; ,
\eqno (4.14)
$$
and
$$
\int_{\Sigma} {\rm Tr} \left[{\tilde p}^{ac}\Omega_{ad}
- {1\over 2} {\tilde p}^{ab}\Omega_{ab}
\; \delta_{d}^{c}\right]
\xi^d \; d^{3}x_{c} \equiv
\int_\Sigma {\widetilde G}^{L}_{\ d} \;
e_{\; \; L}^{c} \; \xi^{d} \; d^{3}x_{c}=0 \; .
\eqno (4.15)
$$
\vskip 1cm
\leftline {\bf 5. - Reduction of constraints to Ashtekar's form.}
\vskip 1cm
In this section we show that the previous equations reproduce
Ashtekar's results for a Palatini Lagrangian [13].
For this purpose,
we choose the {\it adapted} local coordinates defined
by the condition $x^{0}$ = constant, while the remaining three
coordinates are denoted by the spatial indices $i,j,k$. In view
of the covariance of the Lagrangian formalism, it is enough to
prove the equivalence with Ashtekar's constraint analysis
on spacelike hypersurfaces in this
particular coordinate system. The
constraints (4.14)-(4.15) are then found to take the form
(from the arbitrariness of $\lambda^{IJ}$ and $\xi^{d}$)
$$
(D_{a}{\tilde p}^{a0})_{IJ} =0 \; ,
\eqno (5.1)
$$
which corresponds to the Gauss constraint [9,13], and
$$
{\rm Tr} \left[{\tilde p}^{a0}\Omega_{ad}
-{1\over 2} {\tilde p}^{ab}\Omega_{ab}
\; \delta^{0}_{d}\right]=0 \; .
\eqno (5.2)
$$
Note that (5.1) reflects the invariance of general relativity
under local Lorentz transformations, and leads to six
independent constraint equations. They may be further split into
internal rotations and boosts [9,13].
The last four constraints (5.2) can be split into two parts,
for $d \not = 0$ one has
$$
{\rm Tr} \Bigr[{\tilde p}^{a0}\Omega_{aj}\Bigr]
={\rm Tr} \Bigr[{\tilde p}^{i0}\Omega_{ij}\Bigr]=0 \; ,
\eqno (5.3)
$$
which can be identified with the vector constraint;
and for $d=0$
$$ \eqalignno{{\rm Tr} \left[{\tilde p}^{a0}\Omega_{a0}
- {1\over 2} {\tilde p}^{ab}
\Omega_{ab} \right]&
={\rm Tr} \left[{\tilde p}^{i0}\Omega_{i0} -
{1\over 2} {\tilde p}^{i0}\Omega_{i0} -
{1\over 2} {\tilde p}^{0i}\Omega_{0i}-
{1\over 2} {\tilde p}^{ij}\Omega_{ij}\right]\cr
&=- {1\over 2} {\rm Tr} \Bigr[{\tilde p}^{ij}
\Omega_{ij}\Bigr]=0 \; ,
&(5.4)\cr}
$$
which is the Hamiltonian constraint.
To see that (5.4) implies the Hamiltonian
constraint in its usual form, we point out that from (3.5)
$$
{\rm Tr} \Bigr[{\tilde p}^{ij}\Omega_{ij}\Bigr]=
2 e^{-1} (g^{00})^{-1} {\rm Tr} \Bigr[
{\tilde p}^{i0} {\tilde p}^{j0}\Omega_{ij}\Bigr]
+2(g^{00})^{-1}g^{j0}
{\rm Tr} \Bigr[{\tilde p}^{i0}\Omega_{ij}\Bigr] \; .
\eqno (5.5)
$$
Then from (5.3)-(5.5) it follows that
$$
{\rm Tr} \Bigr[{\tilde p}^{i0}{\tilde p}^{j0}
\Omega_{ij}\Bigr]=0 \; ,
\eqno (5.6)
$$
which is the familiar constraint equation quadratic in
the momenta [13].
\vskip 1cm
\leftline{\bf 6. - Preservation of constraints.}
\vskip 1cm
We have now to prove explicitly that the constraints
(4.14)-(4.15) are preserved in our multisymplectic approach.
For this purpose, we begin by requiring that the
following Lie derivative:
$$
{\cal I} \equiv \int_{\Sigma}L_{\eta} \biggr[{\rm Tr}
\Bigr({\tilde p}^{ac}\Omega_{ad}-{1\over 2}
{\tilde p}^{ab}\Omega_{ab} \; \delta_{d}^{c}\Bigr)
\xi^{d}\biggr] \; d^{3}x_{c} \; ,
\eqno (6.1)
$$
should vanish, where $L_{\eta}$ denotes the Lie derivative
along a smooth vector field $\eta$. The integral (6.1)
results from taking differences of the constraints evaluated
at two generic spacelike hypersurfaces $\Sigma$ and $\Sigma'$,
related by a one-parameter flow generated by $\eta$.
Thus, since the
Lie derivative along $\eta$ of a weight-$w$ vector density
$\widetilde V$ reads [26]
$$
\Bigr(L_{\eta}{\widetilde V}\Bigr)^{a}
=\partial_{b}\Bigr({\widetilde V}^{a}\eta^{b}
-{\widetilde V}^{b} \eta^{a}\Bigr)
+ \eta^{a} \partial_{b}{\widetilde V}^{b}
+(w-1) {\widetilde V}^{a}\partial_{b}\eta^{b} \; ,
\eqno (6.2)
$$
one finds that (6.1) takes the form
$$
{\cal I}=2\int_{\partial \Sigma}
{\widetilde G}_{\; \; d}^{c}
\; \xi^{d} \eta^{b} \; d^{2}x_{bc}
+\int_{\Sigma}\eta^{a} {\widetilde G}_{\; \; c}^{b}
\; \nabla_{b}\xi^{c} \; d^{3}x_{a}
+ \int_{\Sigma} \eta^{a} \; \xi^{c} \;
\nabla_{b} {\widetilde G}_{\; \; c}^{b}
\; d^{3}x_{a} \; ,
\eqno (6.3)
$$
where ${\widetilde G}_{\; \; b}^{a} \equiv e \; G_{\; \; b}^{L}
\; e_{\; \; L}^{a}$ (see (2.16)),
and $\nabla$ is the (torsion-free) connection which annihilates
the tetrad. Hence
$\nabla_{a}{\widetilde V}^{a}=\partial_{a}{\widetilde V}^{a}$ [26].
The first term in (6.3)
vanishes by virtue of the boundary
conditions imposed in sect. {\bf 4},
and the second integral vanishes along solutions of the field
equations. Last, the third integral vanishes by virtue of the
contracted Bianchi identities with respect to the space-time
connection $\nabla$.

Similarly, one finds that also the Gauss-law constraint
(4.14) is preserved, since
$$ \eqalignno{
\; & \int_{\Sigma}L_{\eta}\biggr[\lambda^{IJ}
\Bigr(D_{a}{\tilde p}^{ac}\Bigr)_{IJ}\biggr] \; d^{3}x_{c}
=2\int_{\partial \Sigma}
\eta^{b}\lambda^{IJ}
\Bigr(D_{a}{\tilde p}^{ac}\Bigr)_{IJ} \; d^{2}x_{bc} \cr
&+\int_{\Sigma}\eta^{c} \Bigr(\partial_{b}\lambda^{IJ}\Bigr)
\Bigr(D_{a}{\tilde p}^{ab}\Bigr)_{IJ} \; d^{3}x_{c}
+\int_{\Sigma}\eta^{c} \; \lambda^{IJ} \;
\partial_{b} \; \Bigr(D_{a}{\tilde p}^{ab}\Bigr)_{IJ}
\; d^{3}x_{c} \; .
&(6.4)\cr}
$$
Again, the boundary term vanishes by virtue of the boundary
conditions of sect. {\bf 4}, and the second term vanishes
on imposing the field equations. Moreover, the third term
on the right-hand side of (6.4) vanishes by virtue of the
identity
$$
\partial_{b} \Bigr(D_{a}{\tilde p}^{ab}\Bigr)_{IJ}
={1\over 2} \Bigr[\Omega_{ba},
{\tilde p}^{ab}\Bigr]_{IJ}=0 \; .
\eqno (6.5)
$$

Another way to derive these results is to consider the
four-dimensional volume integral of
$\nabla_{c} \Bigr({\widetilde G}_{\; \; d}^{c} \;
\xi^{d} \Bigr)$, taken over a region $V$ whose boundary
is the disjoint union of two hypersurfaces $\Sigma$
and $\Sigma'$. On applying the Leibniz rule and imposing
the field equations, one finds the equation
$$
\int_{\Sigma'}{\widetilde G}_{\; \; d}^{c}
\; \xi^{d} \; d^{3}x_{c}
-\int_{\Sigma}{\widetilde G}_{\; \; d}^{c}
\; \xi^{d} \; d^{3}x_{c}
=\int_{V} \xi^{d} \; \nabla_{c}
{\widetilde G}_{\; \; d}^{c} \; d^{4}x \; .
\eqno (6.6)
$$
Again, the preservation of constraints is achieved by virtue
of the contracted Bianchi identities. A similar argument
can be applied to prove the preservation of (4.14)
(cf. (6.4)).

In a Palatini formalism, one has also to
consider the extra constraints corresponding to second-class
constraints [13]. In the Hamiltonian formulation, second-class
constraints arise since one starts from tetrads and connection
one-forms, while the constraint analysis makes it convenient
to replace the tetrads by the momenta. To prove equivalence
between these two formulations it is then necessary to restrict
the momenta so as to recover general relativity. One then
finds six primary second-class constraints and six
secondary second-class constraints [13]. This leads to the
two degrees of freedom of real general relativity (page 59
of ref. [13]). A similar problem occurs in our analysis,
if the action (2.1) is re-expressed in terms of
the multimomenta. One has then to consider additional
conditions obeyed by the multimomenta.

In the light of (3.1) and (3.3), the second-class constraint
equations found in ref. [13] take the form
$$
n_{c} \; n_{d} \; \epsilon^{IJKL} \;
{\tilde p}_{\; \; \; IJ}^{ac} \;
{\tilde p}_{\; \; \; KL}^{bd}=0 \; ,
\eqno (6.7)
$$
$$
n_{d} \; n_{f} \; \epsilon^{IJKL} \;
{\tilde p}_{\; \; \; I}^{cd \; \; \; M}
\biggr[{\tilde p}_{\; \; \; MJ}^{af} \;
D_{c}\Bigr(n_{h}{\tilde p}^{bh}\Bigr)_{KL}
+{\tilde p}_{\; \; \; MJ}^{bf} \;
D_{c}\Bigr(n_{h}{\tilde p}^{ah}\Bigr)_{KL}\biggr]=0
\; ,
\eqno (6.8)
$$
where $n^{a}$ is the unit timelike normal to the spacelike
hypersurface $\Sigma$. Note that our constraint
equations (4.14)-(4.15) and (6.7)-(6.8) are all expressed
in terms of the multimomenta.

At a Lagrangian level, however, there are no primary constraints
[4,20], since one deals with the pull-back on a manifold
corresponding to the primary-constraint submanifold of
the Hamiltonian formalism.
The problem remains to derive the constraints (if any)
{\it corresponding} to the secondary second-class constraints
of the Hamiltonian formalism (of course, constraints cannot
be divided into first- and second-class in a Lagrangian
framework, since no Poisson brackets exist).
They cannot be derived from
the multimomentum map, which only applies to the analysis
of constraints which are a counterpart
of first-class constraints. In ref. [20], we were able to
use the Gotay-Nester Lagrangian analysis in a
finite-dimensional model to derive the Lagrangian counterpart
of secondary second-class constraints. However,
we do not yet know whether
that technique can be extended to our
formulation of general relativity.

For this purpose,
we are currently investigating a more general set of
equations obeyed by the multimomenta, i.e. (cf. [26,27])
$$
\epsilon_{abcd} \; {\tilde p}_{\; \; \; IJ}^{ab}
\; {\tilde p}_{\; \; \; KL}^{cd}={\widetilde T}_{[IJKL]} \; ,
\eqno (6.9)
$$
where ${\widetilde T}$ is a tensor density proportional
to $\epsilon_{IJKL}$.
Eq. (6.9) admits, as a particular case,
the condition of simplicity of the multimomenta
(e.g. set $I=K,J=L$ in (6.9)) in the abstract indices $a,b$,
as well as the analogous of (6.7) where the roles of
space-time and internal indices are interchanged,
i.e.  $n^{J} \; n^{L} \; {\widetilde T}_{[IJKL]}=0$, where
$n^{J} \equiv e_{a}^{\; \; J} \; n^{a}$.
\vskip 10cm
\leftline {\bf 7. - Self-dual gravity.}
\vskip 1cm
The Hamiltonian formulation of self-dual gravity has received
careful consideration in the current literature [13]. The
corresponding formalism, however, is not manifestly covariant.
To overcome this problem, Samuel [28], and, independently,
Jacobson and Smolin [29,30], proposed a Lagrangian
approach based on a self-dual action.
The key idea is to take complex (co)tetrads on a real Lorentzian
four-manifold [31], and then express the action functional
in terms of the tetrad and of the self-dual part of the
connection:
$$
{ }^{+}\omega \equiv {1\over 2} \Bigr(\omega - i { }^{*}\omega
\Bigr) \; .
\eqno (7.1)
$$
Remarkably, the complex self-dual action
$$
S_{SD} \equiv {1\over 2} \int_{M} d^{4}x \;
e \; e^{aI} \; e^{bJ} \; \Omega_{abIJ}
({ }^{+}\omega)
\eqno (7.2)
$$
is related to the real Palatini action (2.1) by [32]
$$
S_{SD}[e,{ }^{+}\omega(e)]={1\over 2}S_{P}[e,\omega(e)]
-{i\over 8} \int_{M}d^{4}x \; e \; e^{aI} \;
\epsilon_{I}^{\; \; \; JKL} \;
\Omega_{aJKL}(\omega(e)) \; ,
\eqno (7.3)
$$
where the second term in (7.3) vanishes for vanishing torsion,
since then $\Omega_{a[JKL]}=0$. Hence the resulting field
equations for real general relativity
are equivalent, and the corresponding constraints
are first-class only (before imposing reality conditions)
and polynomial (page 64 of ref. [13]).

The self-dual equations are
hence obtained by replacing the full connection with the self-dual
connection in secs. {\bf 2}, {\bf 4},
{\bf 5} and {\bf 6}. Thus, on defining
$$
{ }^{+}\sigma^{ac} \equiv
{ }^{+}{\tilde p}_{\; \; \; IJ}^{ac}
\; \lambda^{IJ} \; ,
\eqno (7.4)
$$
$$
{ }^{+}\rho^{ac} \equiv
{ }^{+}{\tilde p}_{\; \; \; IJ}^{ac}
\; { }^{+}\omega_{b}^{\; \; IJ} \; \xi^{b} \; ,
\eqno (7.5)
$$
the constraint equations become (cf. (4.12)-(4.13))
$$
\int_{\Sigma}\partial_{a}{ }^{+}\sigma^{ac} \; d^{3}x_{c}
-\int_{\Sigma} \lambda^{IJ}
(D_{a}{ }^{+}{\tilde p}^{ac})_{IJ}\; d^{3}x_{c}=0 \; ,
\eqno (7.6)
$$
$$
\int_{\Sigma}\partial_{a}{ }^{+}\rho^{ac} \; d^{3}x_{c}
-\int_\Sigma {\rm Tr} \left[ { }^{+}{\tilde p}^{ac}
\Omega_{ad}({ }^{+}\omega)
- {1\over 2} { }^{+}{\tilde p}^{ab}\Omega_{ab}({ }^{+}\omega)
\; \delta_{d}^{c}\right]
\xi^d \; d^{3}x_{c} =0 \; .
\eqno (7.7)
$$
In other words, in the most recent presentations [13,31], the
equivalence between general relativity with a Palatini action
and its self-dual version is proved. In this paper we follow
the same argument and we find entirely analogous results, while
the multisymplectic framework enables one to preserve covariance.
Moreover, no extra constraints are necessary to recover the
original content of the theory, when the action (7.2) is
re-expressed in terms of the multimomenta (cf. [13]).
As a last step, one has to impose suitable
reality conditions to recover real general relativity [13,31].
Their formulation in the multisymplectic framework is not
studied in our paper, and is a subject for further research.

We think one should emphasize again that, in the canonical-gravity
approach to self-duality, one takes complex (co)tetrads on real
space-time four-manifolds. By contrast, in other branches of
modern relativity [33], one is interested in
four-complex-dimensional complex-Riemannian manifolds. In such a
case, no complex conjugation can be defined, since this map is not
invariant under holomorphic coordinate transformations, and no
four-real-dimensional sub-manifold can
in general be singled out [33]. Hence the problem of reality
conditions to recover real general relativity cannot even be
addressed in the complex-Riemannian framework. The corresponding
theory of self-duality involves the Weyl spinors, and is not
equivalent to the model outlined in this section [33].
\vskip 1cm
\leftline {\bf 8. - Results and open problems.}
\vskip 1cm
This paper has studied Palatini and self-dual gravity theories
by using tetrad formalism and multisymplectic techniques
in Lorentzian four-manifolds (cf. [34]).
Our results are as follows.
\vskip 0.3cm
\noindent
(i) The first-class constraint equations of Palatini theory
are given by (4.12)-(4.13).
Interestingly, boundary terms occur, and they vanish
under the sufficient conditions listed
at the end of sect. {\bf 4}.
\vskip 0.3cm
\noindent
(ii) The Hamiltonian constraint of general relativity
is linear in the multimomenta, as shown in (5.3)-(5.5).
Indeed, on studying Palatini formalism, Eq. (3) of chapter 4
of ref. [13] implicitly expresses this property. Our analysis,
however, makes it more evident. The multisymplectic
framework of sect. {\bf 2}, and the multimomenta formalism
of the following sections, seem to add evidence in favour
of the Lagrangian point of view being able to
supplement the Hamiltonian formalism to get a better understanding
of the gravitational field. Moreover, the linearity in
the multimomenta of all constraint equations may have far
reaching consequences for the quantization program.

It should be emphasized that our analysis has not improved
the understanding of the initial-value problem in general
relativity. However, our equations are covariant in that
they do not depend on a particular spacelike or null
hypersurface, and they naturally lead to study the space
of multimomenta. Hence
they are a step towards a covariant formulation
of relativistic theories of gravitation regarded as
constrained systems. Within this framework, it may be
interesting to analyze 2+1 gravity and theories with
non-vanishing torsion.

A naturally occurring question is
whether our classical analysis can be used to
formulate an approach to non-perturbative quantum gravity
which improves the results obtained within
the more familiar Hamiltonian framework [13].
For this purpose, it appears necessary to get a better understanding
of covariant Poisson brackets [35], and possibly of the
geometric quantization program [36,37]. Hence we do not yet know
the key features of the resulting quantum theory.
However, the elegance of the mathematical formalism, and its wide
range of applications at the classical level, make us feel
that new perspectives in non-perturbative quantum gravity
are in sight.
\vskip 10cm
\leftline {\bf Appendix.}
\vskip 0.3cm
Since our paper is primarily addressed to physicists interested
in general relativity, we limit ourselves to a very brief
outline of some geometric ideas used in our investigation.
An extended treatment may be found in ref. [23].

In our paper, the notation $J^{1}(Y)$ means what follows [24].
Let $X$ be a manifold and let
$Y$ be a fibre bundle having $X$ as its base space, with
projection map $\pi_{XY}$. A fibre is then given by
$\pi_{XY}^{-1}(x)$, where $x \; \in \; X$. Moreover, let
$\gamma: T_{x}X \rightarrow T_{y}Y$ be a linear map between
the tangent space to $X$ at $x$ and the tangent space to
$Y$ at $y \; \in \; \pi_{XY}^{-1}(x)$. Now, given a point $y$
belonging to the fibre $Y_{x}$ through $x \; \in \; X$, we
consider all $\gamma$ maps relative to $y \; \in \; Y_{x}$.
This leads to a fibre bundle $J^{1}(Y)$ having the fibre bundle
$Y$ as its base space and fibres given by the $\gamma$ maps.
Such a $J^{1}(Y)$ is called the {\it one-jet bundle} on $Y$.
If $\varphi^{(A)}(x^{\mu})$ is a section of $Y$, the
{\it tangent lift} of $\varphi^{(A)}$ to a section of $J^{1}(Y)$
is denoted by $j^{(1)}(\varphi)$. It is given by the map
$$
j^{(1)}:\varphi \rightarrow \biggr(x^{\mu},
\varphi^{(A)}(x^{\mu}),
{\partial \varphi^{(A)}(x^{\mu})
\over \partial x^{\nu}}\biggr) \; .
\eqno (A.1)
$$

The {\it Legendre map}, whose pull-back
appears in Eq. (4.1), is a function
$$
{\cal FL}: J^{1}(Y) \rightarrow \Bigr[J^{1}(Y)\Bigr]^{*} \; ,
\eqno (A.2)
$$
where $\Bigr[J^{1}(Y)\Bigr]^{*}$ is a fibre bundle having
$Y$ as its base space, and whose fibre through $y \in Y$ is
given by the affine maps on the elements of the fibre of
$J^{1}(Y)$, with coefficients in the bundle of $n+1$ forms
on $X$ at $x$. Hence $\Bigr[J^{1}(Y)\Bigr]^{*}$
is called the dual of the one-jet bundle $J^{1}(Y)$.
If, in local coordinates, $J^{1}(Y)$ is described by
$\Bigr(x^{\mu},y^{(A)},v_{\; \; \; \; \mu}^{(A)}\Bigr)$, the
expression of its dual in local coordinates is given by
$\Bigr(x^{\mu},y^{(A)},p,p_{(A)}^{\; \; \; \; \mu}\Bigr)$,
where the definitions of momenta and of Legendre map yield
$$
p_{(A)}^{\; \; \; \; \mu} \equiv
{\partial L \over \partial v_{\; \; \; \; \mu}^{(A)}}
\; ,
\eqno (A.3)
$$
$$
p \equiv L - p_{(A)}^{\; \; \; \; \mu} \;
v_{\; \; \; \; \mu}^{(A)} \; .
\eqno (A.4)
$$
\vskip 0.3cm
\centerline {$* \; * \; *$}
\vskip 0.3cm
The authors are much indebted to Giuseppe Marmo for encouraging
their work and teaching them all what they know about
Lagrangian field theory. Our research was supported in part by
the European Union under the Human Capital and Mobility Program.
Gabriele Gionti
is grateful to the military authorities of the Military District
of Trieste, and in particular to the Head of the Military District,
Col. Luciano Monaco, for making it possible for him to conclude
this work during his military service.
\vskip 0.3cm
\leftline {\it REFERENCES}
\vskip 0.3cm
\item {[1]}
P. A. M. DIRAC: {\it Lectures on Quantum Mechanics} (Yeshiva
University, New York, 1964).

\item {[2]}
P. A. M. DIRAC: {\it Lectures on Quantum Field Theory},
Belfer Graduate School of Science, Monographs Series,
Number Three (Yeshiva University, New York, 1966).

\item {[3]}
A. HANSON, T. REGGE and C. TEITELBOIM: {\it Constrained
Hamiltonian Systems} (Accademia dei Lincei, Rome, 1976).

\item {[4]}
G. MARMO, N. MUKUNDA and J. SAMUEL: {\it Riv. Nuovo Cimento},
{\bf 6}, 1 (1983).

\item {[5]}
G. ESPOSITO: {\it Quantum Gravity, Quantum Cosmology and
Lorentzian Geometries}, Lecture Notes in Physics, New
Series m: Monographs, Vol. m12, Second Corrected and Enlarged
Edition (Springer, Berlin, 1994).

\item {[6]}
B. S. DE WITT: {\it Phys. Rev.}, {\bf 160}, 1113 (1967).

\item {[7]}
A. ASHTEKAR: {\it Phys. Rev. Lett.}, {\bf 57}, 2244 (1986).

\item {[8]}
A. ASHTEKAR: {\it Phys. Rev. D}, {\bf 36}, 1587 (1987).

\item {[9]}
A. ASHTEKAR: {\it New Perspectives in Canonical Gravity}
(Bibliopolis, Naples, 1988).

\item {[10]}
C. ROVELLI and L. SMOLIN: {\it Phys. Rev. Lett.}, {\bf 61},
1155 (1988).

\item {[11]}
C. ROVELLI and L. SMOLIN: {\it Nucl. Phys. B}, {\bf 331}, 80 (1990).

\item {[12]}
A. ASHTEKAR, C. ROVELLI and L. SMOLIN:
{\it Phys. Rev. D}, {\bf 44}, 1740 (1991).

\item {[13]}
A. ASHTEKAR: {\it Lectures on Non-Perturbative Canonical
Gravity} (World Scientific, Singapore, 1991).

\item {[14]}
J. N. GOLDBERG, J. LEWANDOWSKI and C. STORNAIOLO:
{\it Commun. Math. Phys.}, {\bf 148}, 377 (1992).

\item {[15]}
A. ASHTEKAR and C. J. ISHAM: {\it Class. Quantum Grav.}, {\bf 9},
1433 (1992).

\item {[16]}
J. SNIATYCKI: {\it On the canonical formulation of general
relativity}, in {\it Proc. Journ\'ees Relativistes}
(Facult\'e des Sciences, Caen, 1970).

\item {[17]}
W. SZCZYRBA: {\it Commun. Math. Phys.}, {\bf 51}, 163 (1976).

\item {[18]}
J. NOVOTNY: {\it On the geometric foundations of the Lagrange
formulation of general relativity},
in {\it Differential Geometry}, eds. Gy. Soos
and J. Szenthe, Eds. Colloq. Math. Soc. J. Bolyai
{\bf 31} (North-Holland, Amsterdam, 1982).

\item {[19]}
S. W. HAWKING: {\it The path-integral approach to quantum
gravity}, in {\it General Relativity, an Einstein Centenary
Survey}, eds. S. W. Hawking and W. Israel (Cambridge University
Press, Cambridge, 1979).

\item {[20]}
G. ESPOSITO, G. GIONTI, G. MARMO and C. STORNAIOLO:
{\it Nuovo Cimento B}, {\bf 109}, 1259 (1994).

\item {[21]}
G. MENDELLA, G. MARMO and W. M. TULCZYJEW: {\it J. Phys. A},
{\bf 28}, 149 (1995).

\item {[22]}
L. SMOLIN: {\it Class. Quantum Grav.}, {\bf 9}, 883 (1992).

\item {[23]}
M. J. GOTAY, J. ISENBERG, J. E. MARSDEN and R. MONTGOMERY:
{\it Momentum Mappings and the Hamiltonian Structure of
Classical Field Theories with Constraints}
(Springer, Berlin, in press).

\item {[24]}
G. GIONTI: {\it Constraints' Theory and General Relativity},
Thesis, University of Napoli (1993).

\item {[25]}
J. N. GOLDBERG: {\it Found. Phys.}, {\bf 14}, 1211 (1984).

\item {[26]}
J. A. SCHOUTEN: {\it Ricci-Calculus} (Springer, Berlin, 1954).

\item {[27]}
R. L. BISHOP and S. I. GOLDBERG: {\it Tensor Analysis on
Manifolds} (Dover, New York, 1980); R. CAPOVILLA, J. DELL
and T. JACOBSON: {\it Class. Quantum Grav.}, {\bf 8},
59 (1991).

\item {[28]}
J. SAMUEL: {\it Pramana J. Phys.}, {\bf 28}, L429 (1987).

\item {[29]}
T. JACOBSON and L. SMOLIN: {\it Phys. Lett. B}, {\bf 196}, 39 (1987).

\item {[30]}
T. JACOBSON and L. SMOLIN: {\it Class. Quantum Grav.}, {\bf 5},
583 (1988).

\item {[31]}
J. ROMANO: {\it Gen. Rel. Grav.}, {\bf 25}, 759 (1993).

\item {[32]}
H. A. MORALES-T\'ECOTL and G. ESPOSITO:
{\it Nuovo Cimento B}, {\bf 109}, 973 (1994).

\item {[33]}
R. PENROSE and W. RINDLER: {\it Spinors and Space-Time, Vol. II}
(Cambridge University Press, Cambridge, 1986); G. ESPOSITO:
{\it Complex General Relativity}, Fundamental Theories of
Physics, Vol. 69 (Kluwer, Dordrecht, 1995).

\item {[34]}
J. SNIATYCKI: {\it Rep. Math. Phys.}, {\bf 19}, 407 (1984).

\item {[35]}
J. E. MARSDEN, R. MONTGOMERY, P. J. MORRISON and W. B. THOMPSON:
{\it Ann. Phys. (N.Y.)}, {\bf 169}, 29 (1986).

\item {[36]}
J. SNIATYCKI: {\it Geometric Quantization and Quantum Mechanics}
(Springer, Berlin, 1980).

\item {[37]}
N. M. J. WOODHOUSE: {\it Geometric Quantization} (Oxford University
Press, Oxford, 1980).

\bye